\documentclass[letterpaper,10pt,final,balancelastpage,floats,aps,twocolumn]{revtex4}
\usepackage{amsmath}
\usepackage{amssymb}
\usepackage{acronym}
\usepackage{graphicx}

\begin{document}
\title{The Radical Character of the Acenes: \\A Density Matrix
  Renormalization Group Study}

\author{Johannes Hachmann, Jonathan J. Dorando, Michael Avil\'{e}s and Garnet Kin-Lic Chan}
\affiliation{Department of Chemistry and Chemical Biology\\Cornell
  University, Ithaca, New York 14853-1301}
\date{\today}
\begin{abstract}
We present a detailed investigation of the acene series using high-level
wavefunction
theory. Our  \textit{ab-initio} Density Matrix Renormalization Group algorithm
 has enabled us to carry out Complete Active Space
 calculations  on the acenes from napthalene to dodecacene
 correlating  the full $\pi$-valence space. While we find that the
 ground-state is a
 singlet for all chain-lengths, examination of several measures of radical character, including the
natural orbitals, effective number of unpaired electrons, and various
correlation functions, suggests that the longer acene ground-states  are
\textit{polyradical} in nature.  
\end{abstract}
\maketitle

\section{Introduction}

\begin{figure}
\centerline{\includegraphics[width=9cm]{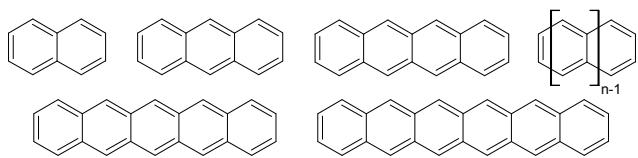}}
\caption{The first few members and the unit cell of the acene series.}
\label{fig:acenes}
\end{figure}

The acenes are the series of ladder-like compounds consisting of
linearly-fused benzene rings (Fig. \ref{fig:acenes}) \cite{Clar1964,Havey1997}.  Due to
their  technological potential \cite{Geerts1998,Dimitrakopoulos2002,Reese2004} and  their intrinsic
value as models for more complex  conjugated
molecules, they have been the subject of many  theoretical and
experimental investigations \cite{BendikovRev,Angliker1982,Kertesz1983,Kivelson1983,Wiberg1997,Houk2001,Bendikov2004,Mondal2006,Bendikov2006}.
In  a number of recent studies, it
has been proposed that longer
acenes may possess an unusual electronic ground-state that is not the
simple closed-shell singlet suggested by molecular orbital arguments.  Based on
extrapolating  the  experimental singlet-triplet gap of the acenes up
to pentacene, Angliker \textit{et al.} \cite{Angliker1982} predicted that the ground state
of higher acenes from nonacene upwards would be  a triplet. 
Density functional calculations by Houk \textit{et al.} \cite{Houk2001} also predicted a
 singlet-triplet cross-over. However, Bendikov \textit{et al.}
 \cite{Bendikov2004} noted that the restricted singlet  density functional ground-state  would become
unstable to an open-shell singlet, or
 \textit{singlet diradical}, configuration for acenes longer than hexacene. The
open-shell singlet-triplet gap for the longest acene studied
(decacene) was estimated as ranging from 1.5 (BLYP/6-31G(d))
to 5.7 (B3LYP/6-31G(d)) kcal/mol.

Despite these intriguing findings, the density functional results
leave many  interesting questions unanswered. For example,  how
diradicaloid are the acenes really compared to conventional diradical systems?
As we go to longer
acenes, might we expect to find tri- and even higher polyradical
ground states? And if so, how do we understand the electronic
structure and bonding in these  states? Such questions, which probe the
essential \textit{many-electron} character of di- (and indeed poly-)
radicalism, are not easily answered  through  density
functional theory based on a single Kohn-Sham determinant.

For this reason we have decided to explore the nature of the acene
ground-state using high-level wavefunction-based electronic structure
theory. The many-electron correlations in radical wavefunctions tell us about the  coupled simultaneous motions of the electrons.
Conceptually, singlet states with unpaired electrons require  multi-configurational
wavefunctions \cite{BallyBordenRev, StantonGaussRev, Slipchenko2002,
Salem1972, Borden1977,BordenBook, Rajca1994,Jung2003,Krylov2005} as used, for example, in the CAS (complete active space)
family of methods \cite{Roos1987}. In the acenes, the ideal choice of  active space would
be the complete $\pi$-valence space, i.e. the set of all conjugated $p_z$
orbitals.  However, the exponential cost of traditional CAS methods as a function of the number
of correlated orbitals and electrons  renders  calculations with the
complete $\pi$-valence space impossible for acenes much longer than
napthalene, which already has 10 conjugated orbitals and
electrons. Consequently, earlier CAS calculations could only use an 
incomplete $\pi$-valence space \cite{Kawashima1999,Bendikov2004}.

The Density Matrix Renormalization Group (DMRG) algorithm
\cite{White1992,White1993,WhiteMartin1999} provides a way to overcome the traditional
exponential complexity of CAS methods in long molecules such as the acenes
\cite{Raghu2002a,Raghu2002b}. We have recently developed a   local
\textit{ab-initio} DMRG method that
computes an essentially exact CAS wavefunction with an effort that
scales only quadratically with the length of the
molecule \cite{Hachmann2006}. Consequently,  we can now extend the range of traditional
CAS calculations in long molecules from about 10 orbitals to 100 active
orbitals or more. 
In the current work we apply our  \textit{ab-initio} DMRG algorithm to
the acene series from napthalene (2-acene) to dodecacene (12-acene),
in all cases correlating exactly the complete $\pi$-valence space. 
First we revisit the question of the relative stabilities of the
singlet and triplet states. Then, using  our
correlated wavefunctions, we embark on a detailed study of the radical
nature of these systems. We find intriguingly that the higher acenes
are not only diradicals, but possess increasing polyradical character. By explicit
visualisation of the electron correlation,  we uncover a coupled motion
of the electrons  that gives a new picture of bonding in
 molecules with extended conjugation, showing that even systems such as the acenes
can continue to provide  fertile sources of surprising electronic structure.


\section{Computational Methodology}

In the present context, we can regard the DMRG as an efficient
way to exactly correlate, in the sense of full configuration interaction, the electrons in the active
space. Details of the DMRG
algorithm as implemented in our \textsc{Block} code are given in
Ref. \cite{Chan2002,Chan2004,Hachmann2006}. Active space full
configuration interaction theory is sometimes referred to as CASCI. Recall that
CASCI is the same as the more common CASSCF (complete active space self-consistent field) method \cite{Roos1987} but lacks  the step of
orbital optimisation. Orbital optimisation is possible within the
DMRG but has not been used here. For the molecules in this
work, the DMRG energies are converged to better than 0.1 kcal/mol and would  be identical  to the so-called CASCI energies if it
were possible to compute these in the traditional manner.

DMRG calculations on the acenes were performed at the UB3LYP/6-31G(d)
\cite{LYP1988,Becke1993} optimised singlet and triplet geometries
which were essentially the same as those used by Bendikov \textit{et al.} \cite{Bendikov2004} (see supporting information for details).
These structures have $D_{2h}$ point-group symmetry. The rung C-C bonds are somewhat longer than the ladder C-C bonds and the ladder C-C bonds display increasing bond alternation towards the ends of the chain.
For example in singlet decacene, the rung and ladder C-C bonds were 1.464{\AA} and 1.405{\AA} respectively at the middle of the chain, while the 
difference in successive ladder C-C bonds lengths was 0.058{\AA} at the end of the chain as compared to 0.010{\AA} at the middle.

The active space was
chosen to be the
complete  $\pi$ valence space, consisting of all conjugated carbon $p_z$
orbitals, and  all $\pi$ electrons were correlated. The $\sigma$
electrons were treated within a frozen-core approximation using the
restricted Hartree-Fock orbitals. The calculations used
either the minimal STO-3G basis \cite{sto3g} (up to dodecacene) or Dunning's 
double-$\zeta$ DZ \cite{dz1, dz2} basis (up to hexacene) as indicated. In the case
of the DZ basis, two $p_z$ orbitals per carbon were used to make a
``double'' complete $\pi$-valence space. Thus whereas e.g. the DMRG/STO-3G
calculations for pentacene correspond to a (22, 22) CASCI, the DMRG/DZ
calculations would correspond to a (22, 44) CASCI. \cite{supportingmaterial} 

\section{The singlet-triplet gap}

\begin{figure}
\centerline{\includegraphics[width=9cm]{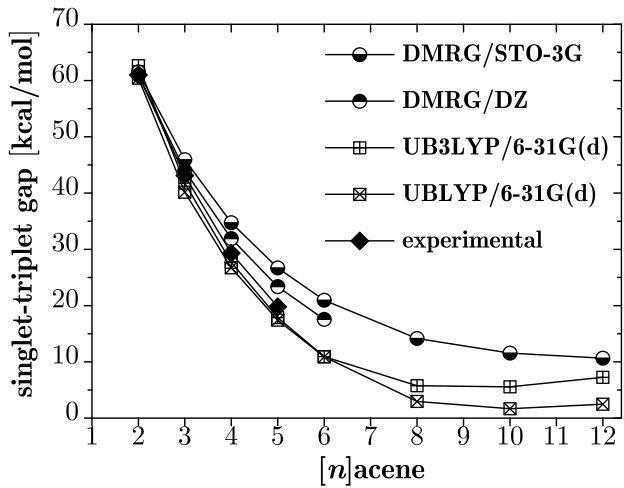}}
\caption  {Singlet-triplet energy gap as a function of the acene
  length. }
\label{fig:stgap}
\end{figure}

In Fig. \ref{fig:stgap} we present the computed DMRG singlet-triplet
energy gaps as a function of the acene length. The calculations on dodecacene 
correspond to  a (50, 50) CASCI and are only made possible through the 
 DMRG algorithm. Included for comparison are the UB3LYP/6-31G(d) and UBLYP/6-31G(d)
 singlet-triplet gaps (using open-shell wavefunctions where stable)
 as first reported by Bendikov \textit{et al.} \cite{Bendikov2004}, which we have recomputed and
extended to the complete set of acenes studied here \cite{g03}.
While experimental triplet energies are somewhat difficult
to compare directly with theoretical gas phase calculations, we
have also included current experimental estimates where available
\cite{BirksBook, Schiedt1997, Sabbatini1982, Burgos1977} \footnote{The hexacene
``experimental'' number
  reported in Houk \textit{et al.} \cite{Houk2001} is in fact the theoretical
extrapolation given in \cite{Angliker1982} and is too low.}.

Our DMRG calculations clearly confirm that the acenes maintain a singlet
ground-state configuration and that there is a finite singlet-triplet
gap for all chain-lengths. Going from the minimal STO-3G to the
DZ basis and the corresponding larger double-active space, the singlet-triplet gap decreases
by a few kcal/mol. With the DZ basis the hexacene DMRG gap
is 17.5 kcal/mol. The remaining error in the DMRG calculations arises
from the neglect of dynamical and $\sigma$-$\pi$ correlations, which
would generally further decrease  the gap size. However, we estimate the effect of
dynamical correlation on the gap to be very small when using the complete (and the double
complete) $\pi$-valence space, on the order of a few kcal/mol. In Table \ref{tab:dynamical_energies} we present
additional CASSCF and CASPT2 results (including the CASSCF and MRMP
calculations of  Kawashima \textit{et al.} \cite{Kawashima1999})  for
the smaller acenes to estimate the effects of dynamical correlation.  CASPT2 \cite{Andersson1992} and MRMP \cite{Hirao1992} both incorporate dynamical correlation on top of
the CASSCF reference through second-order perturbation theory. 
We observe in naphthalene that when using a complete $\pi$-valence  space  the
CASSCF singlet-triplet gap is very close (within 1-2 kcal/mol) to the
CASPT2 singlet-triplet gap. It is only when incomplete active spaces
 are used that the CASPT2/MRMP gap is significantly different from the CASSCF gap. In all cases, the
DMRG complete and double $\pi$-valence space gaps are closer to the experimental result
than the MRMP gap in an incomplete active space. This highlights the
importance of the complete $\pi$-valence space for $\pi$-electron excitations. 


Comparison of the UBLYP and UB3LYP gaps with the experimental
data suggests that the DFT results are  an underestimate. This is
particularly true for UBLYP which substantially underestimates the
gap. Surprisingly, the DFT gaps appear to \textit{increase} between
10-acene and 12-acene.


Using our DMRG data we can extrapolate to the infinite polyacene
limit. Empirically, we find that the singlet-triplet gap is well fitted by an
exponential form $a + be^{-c}$, giving a gap for the infinite chain of 
8.69 $\pm$ 0.95 (STO-3G) and 3.33 $\pm$ 0.39 (DZ) kcal/mol
respectively, somewhat lower than the previous estimate of 12.2
kcal/mol  obtained by  Raghu \textit{et al.}
\cite{Raghu2002b} using the semi-empirical
Pariser-Parr-Pople (PPP) Hamiltonian \cite{PPP1, PPP2}.

\begin{table*}
\begin{center}
\caption{Singlet-triplet gap ($E_\text{triplet} -E_\text{singlet}$) energies in kcal/mol for the
  acene series. }
\label{tab:energies}
\begin{tabular}{c c c c c c}
\hline 
\hline
\text{$[n]$acene} & DMRG/STO-3G & DMRG/DZ & UB3LYP/6-31G(d) &
UBLYP/6-31G(d) & Experiment \\
\hline
2 & 61.5  &    61.0 &  62.6 & 60.4 & 61.0 \cite{BirksBook} \\   
3 & 45.9  &    44.0 &  41.8 & 40.2 & 43.1 \cite{Schiedt1997}\\   
4 & 34.7  &    31.9 &  27.7 & 26.7 & 29.3 \cite{Sabbatini1982}\\   
5 & 26.7  &    23.4 &  17.9 & 17.4 & 19.8 \cite{Burgos1977} \\  
6 & 21.0  &    17.5 &  10.9 & 10.9     \\
8 & 14.2  &         &  5.8  & 3.0 &     \\
10&  11.6 &         &  5.6  & 1.7 &     \\
12&  10.7 &         &  7.3  & 2.5     \\
\hline
\hline
\end{tabular}
\end{center}
\end{table*}

\begin{table}
\begin{center}
\caption{Effect of active-space size and dynamical correlation on
  the singlet-triplet gap in smaller acenes. Complete = complete
  $\pi$-valence space, double = double $\pi$-valence space, partial =
  incomplete active space: 2-acene (8,8), 3- and 4-acenes (12,12). DZP = Dunning DZ basis with polarization
  functions \cite{dz1, dz2} except for results of Kawashima \textit{et al.}
  \cite{Kawashima1999}. All energies in kcal/mol.}
\label{tab:dynamical_energies}
\begin{tabular}{l c c c}
\hline
\hline
\text{$[n]$acene}       & 2          & 3 & 4\\
\hline
complete/DZ             &            &   &  \\
CASSCF                  & 61.1       &   &  \\
CASPT2                  & 60.5       &   & \\
\hline                   
complete/DZP            &            &   &  \\
CASSCF                  & 61.1       &   &  \\
CASPT2                  & 59.7       &   &  \\
\hline
partial/DZP             & \\
CASSCF                  & 67.1       & 60.0\footnotemark[1] & 47.3\footnotemark[1] \\
CASPT2/MRMP             & 56.9       & 46.1\footnotemark[1] & 34.8\footnotemark[1] \\
\hline
complete/STO-3G & \\
DMRG                    & 61.5       & 45.9 & 34.7 \\
double/DZ & \\
DMRG                    & 61.0       & 44.0 & 31.9 \\
\hline

\hline
Expt                    & 61.0  & 43.1 & 29.3 \\
\hline
\hline
\end{tabular}
\footnotetext[1] {CASSCF/MRMP calculations of Kawashima \textit{et al.}
\cite{Kawashima1999}; vertical singlet-triplet gap in a cc-pVDZ basis
without polarization functions on H.} 
\end{center}
\end{table}

\section{Polyradical character of the ground-state}

\label{sec:polyradcharacter}

\begin{figure}
\centerline{\includegraphics[width=9cm]{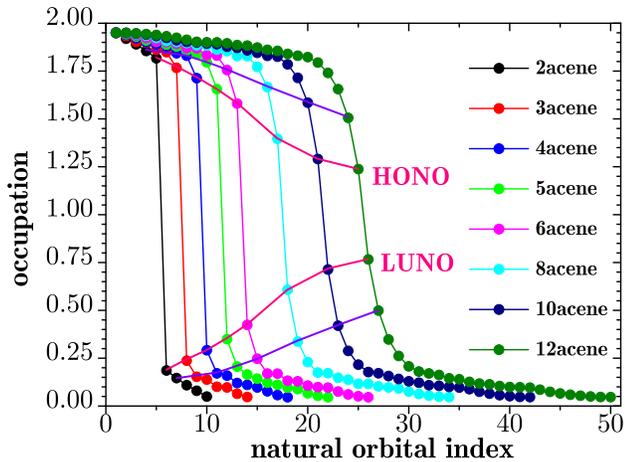}}
\caption{Natural orbital occupation numbers for the acene series in the STO-3G
basis. The lines are guides for the eye to show the evolution of the occupation
  numbers for the near-singly occupied orbitals as a function of chain
  length.}
\label{fig:nocc}
 \end{figure}

\begin{figure*}
\begin{center}
\includegraphics[width=3in]{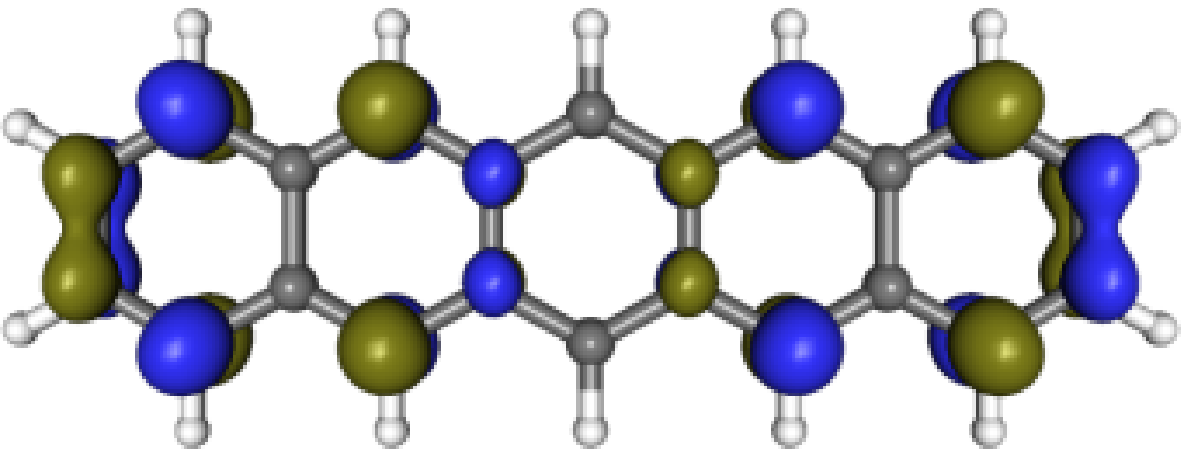} %
\includegraphics[width=3in]{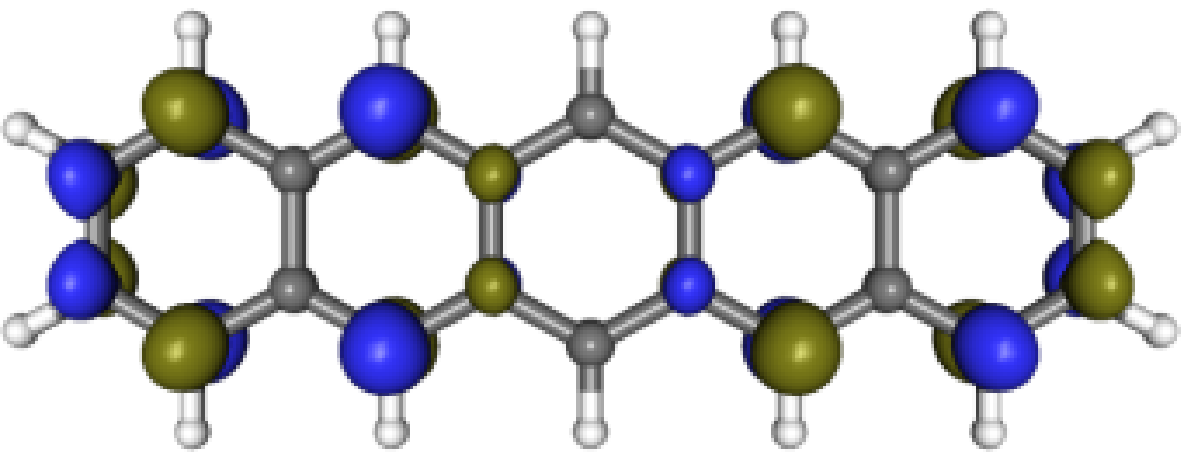} \newline
\includegraphics[width=3in]{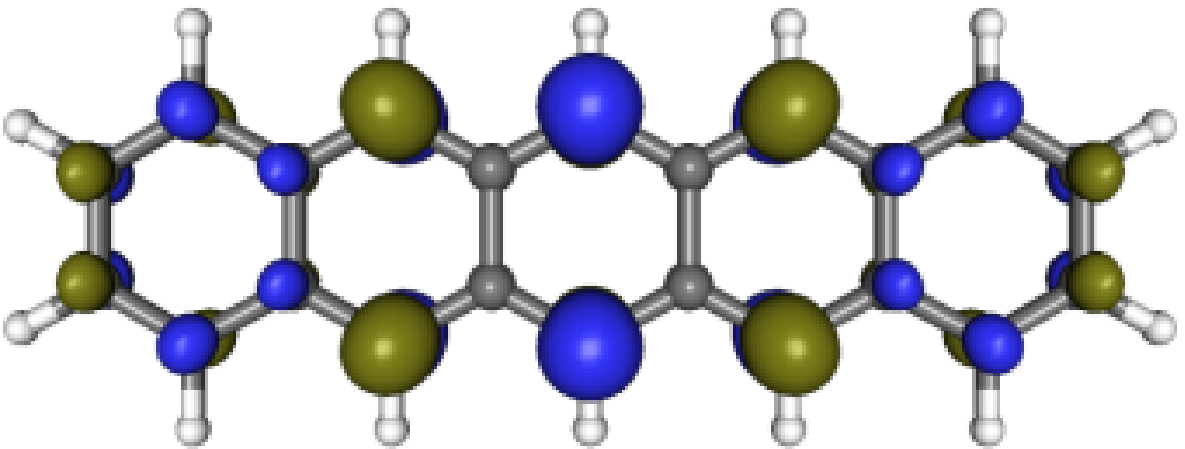} %
\includegraphics[width=3in]{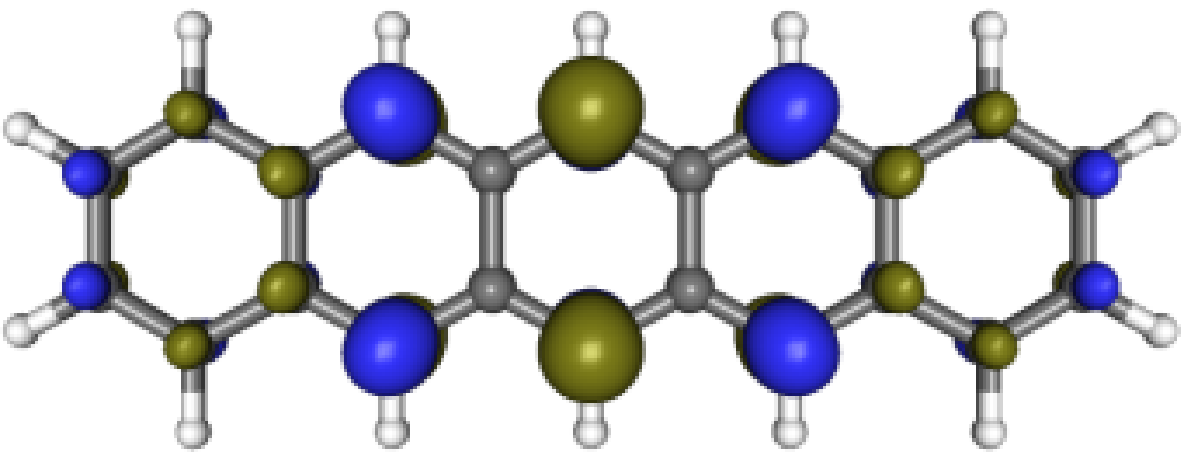} \newline
\includegraphics[width=3in]{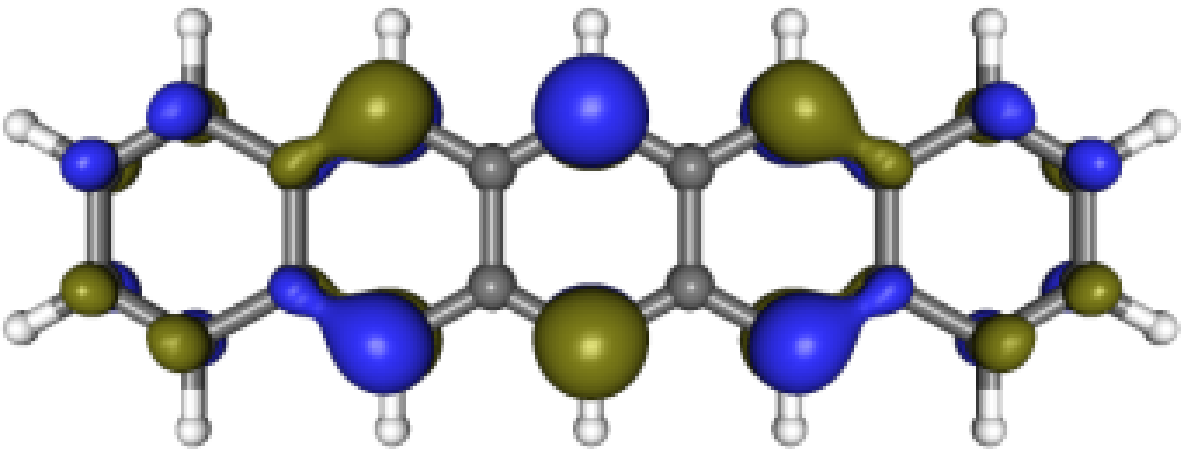} %
\includegraphics[width=3in]{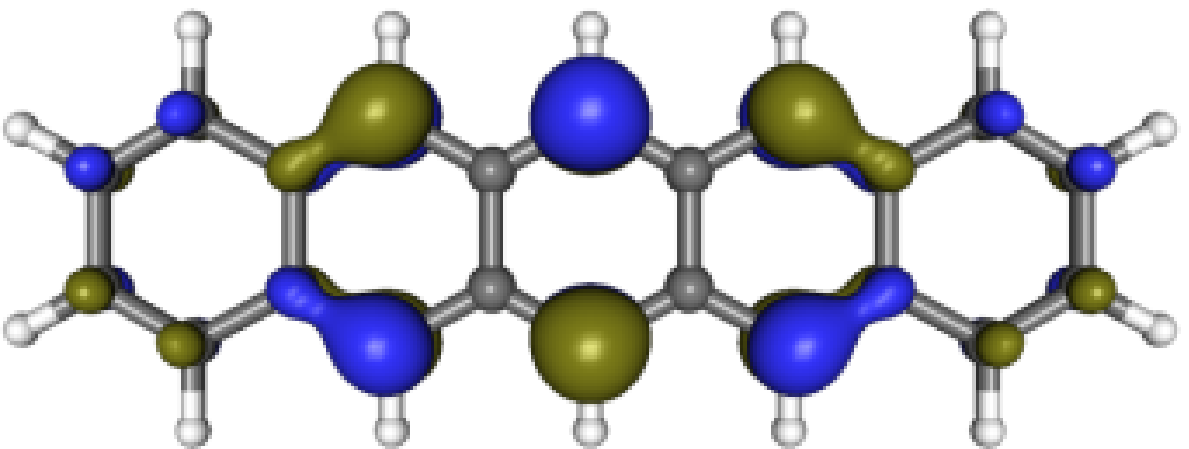} \newline
\includegraphics[width=3in]{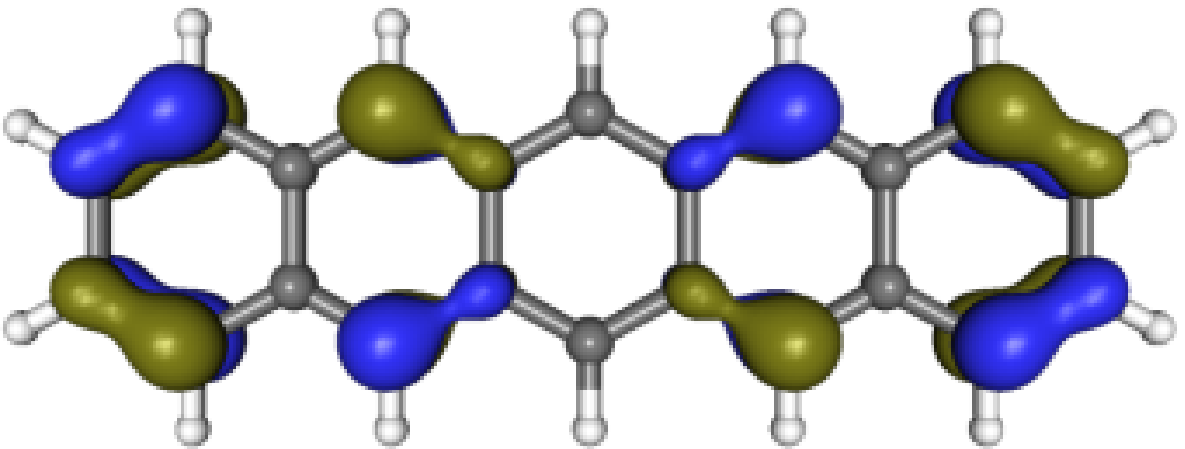} %
\includegraphics[width=3in]{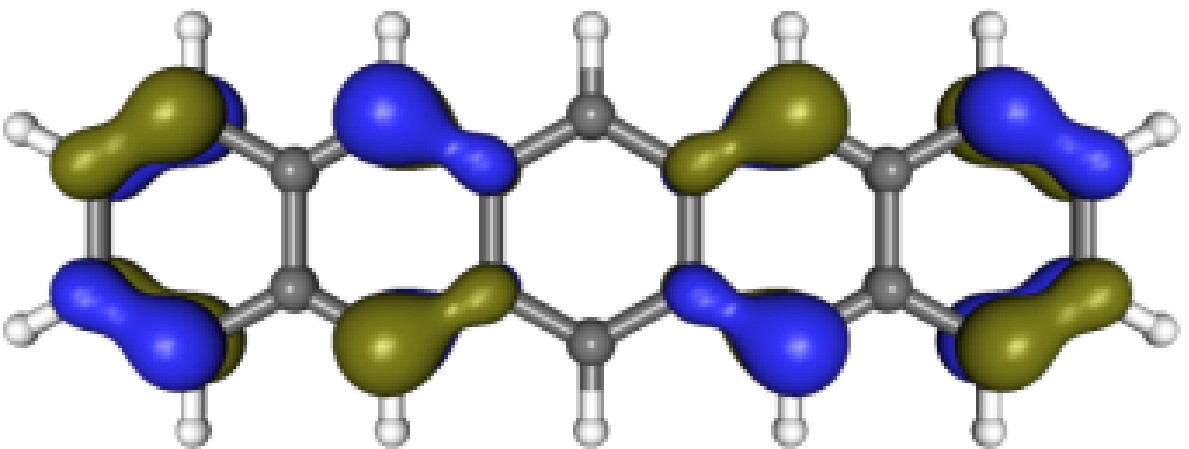}  \newline
\caption{Plots of the molecular orbitals (left: LUMO+1, LUMO, HOMO, HOMO-1) and
natural orbitals (right: LUNO+1, LUNO, HONO, HONO-1) for pentacene.}
\label{fig:orbitalplots}
\end{center}
\end{figure*}

Having established that the acene ground-states are singlets,  are they
then  singlet diradicals as argued by
Bendikov \textit{et al.} \cite{Bendikov2004}? A simple way to establish whether there are unpaired
electrons in a correlated wavefunction is to examine the occupation
numbers of the (spinless) natural orbitals - in a closed shell
configuration, these are always 2 (doubly occupied) or 0 (unoccupied),
while values close to 1 indicate single occupancy and unpaired electrons \cite{Doehnert1980}. 
In
Fig. \ref{fig:nocc} we plot the occupancies of the natural orbitals
for the acene series. We have designated the two orbitals with
occupancies closest to 1 the HONO (``highest occupied natural
orbital'' with occupancy greater than 1) and LUNO (``lowest unoccupied
natural orbital'' with occupancy less than 1)
respectively. These natural orbitals together with usual HOMO and LUMO
are shown in Fig. \ref{fig:orbitalplots}.

\begin{figure}
\centering
\centerline{\includegraphics[width=9cm]{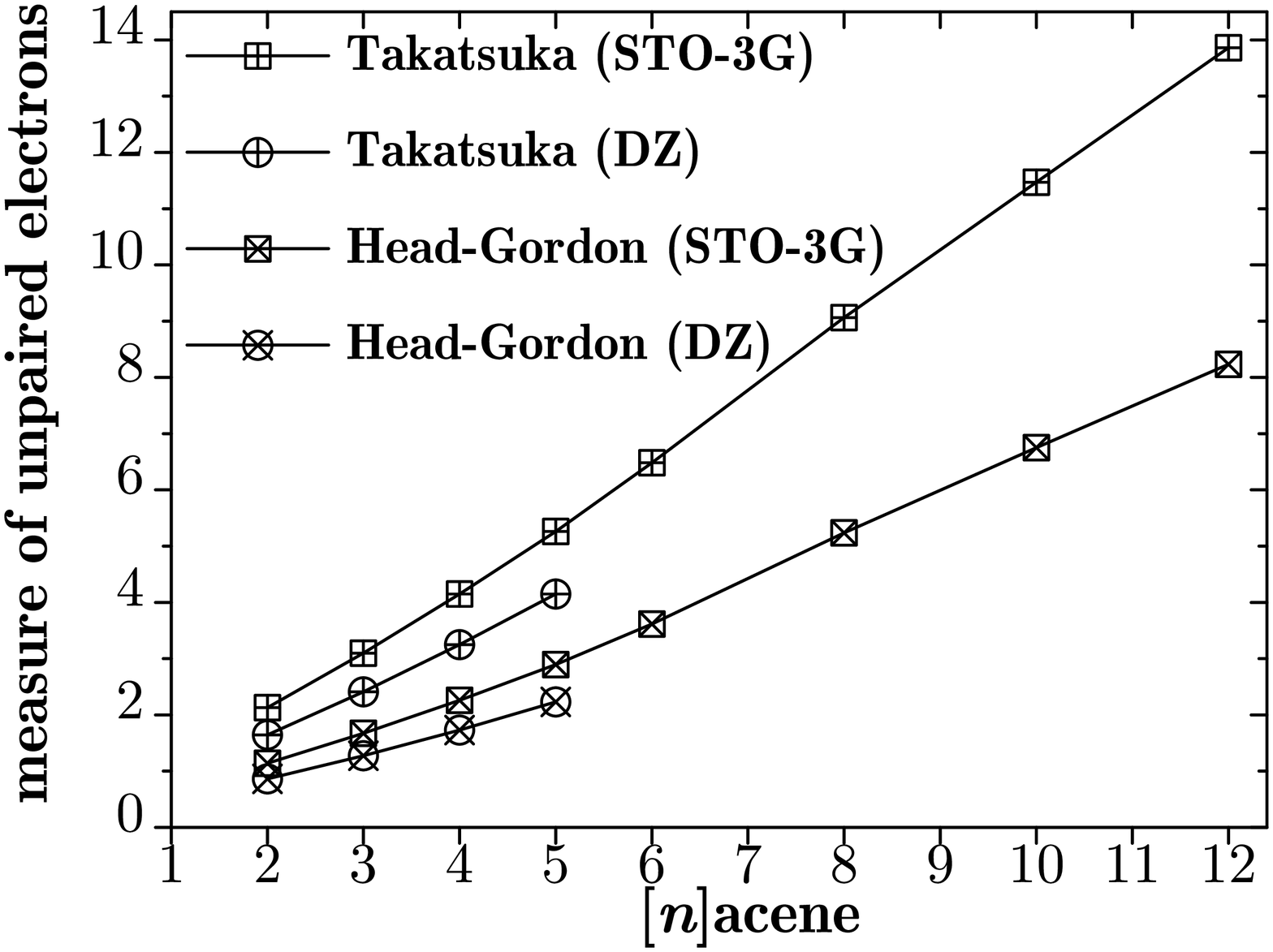}}
\caption{Measures of the number of unpaired electrons in the acene
  series. See comment in text.}
\label{fig:measures}
\end{figure}

As can be seen, as we proceed to longer acenes the
occupancy of the HONO
and LUNO indeed approach  1, which is consistent with
Bendikov \textit{et al.}'s prediction of diradical
character. The DZ basis, while yielding 
 less radical character (e.g. the occupancy of the HONO  in pentacene  is 1.66 and 1.73 using the STO-3G and DZ bases
respectively) shows the same general behaviour (The decreased radical character in the DZ basis is consistent with
the general observation that radical character is reduced by dynamic correlation). However, what is surprising
is that if we follow the \textit{trend} for the next nearest single
occupancy orbitals (the HONO-1 and the LUNO+1), the rate at which they
approach single occupancy is comparable to that of the HONO
and LUNO.  This suggests that if we were to proceed to acenes longer
than the 12-acene, \textit{we would eventually find not  a diradical ground-state,
but a polyradical ground-state}.

Several different measures of the number of ``effectively unpaired'' electrons in
a molecule have previously been proposed. 
While such integrated measures  must contain less information than the underlying distribution of natural
 orbital occupations  examined above, we include them here for completeness.
We have investigated two measures, due to 
Takatsuka \cite{Takatsuka1978a, Takatsuka1978b,
  Bochicchio1998, Staroverov2000a, Staroverov2000b} and
Head-Gordon respectively,
\cite{HeadGordon2003a,Bochicchio2003,HeadGordon2003b}, defined via
\begin{eqnarray}
N_\text{unpaired}^\text{Takatsuka} &= \sum_i 2n_i -n_i^2 \\
N_\text{unpaired}^\text{Head-Gordon} &= \sum_i \min(n_i, 2-n_i)
\end{eqnarray}
Here $n_i$ is the occupation number of the $i$th natural orbital,
which ranges from 0 to 2. The contribution from each orbital is a
maximum when $n_i=1$, whence each orbital contributes 1 electron to
the effective number of unpaired electrons. 
These measures are plotted for the acenes in Fig. \ref{fig:measures}.

Some care must be taken when interpreting Fig. \ref{fig:measures}. Certainly,
the values must not be taken literally; 12-acene does not contain 14
unpaired electrons! Both measures are extensive, meaning that they
increase with the size of the molecule. (This also means that a large enough assembly of nearly closed-shell molecules would
appear to have a substantial number of unpaired electrons using these
measures. In such a case, however, the HONO and LUNO occupation
numbers would \textit{not} change and would stay near 2 and 0 as the number of molecules is increased,
unlike what we see in the acenes). However,   extensive scaling
should only be observed for system
sizes  larger than the typical size of an unpaired
electron. Examination of the gradients of the plots in Fig. \ref{fig:measures} shows an onset of
extensive scaling around hexacene, which is consistent with the observation
of Bendikov \textit{et al.} that the first symmetry breaking of the
density functional calculations occurs also at this chain-length. This
further suggests that we can roughly associate \textit{one unpaired spin with every five to six rings}.

\section{Visualising electron correlations}

\subsection{Theoretical background}

Let us then regard the ground-state of the longer acenes as being a
singlet polyradical. How are we to understand its electronic structure?
We can visualise the simultaneous behaviour of
the multiple electrons involved in the polyradical behaviour through their
correlation functions. Correlation functions have long been used to
understand  bond-alternation and electron correlation in conjugated
systems \cite{Fano1998,Raghu2002a}. We have evaluated three correlation
 functions, the particle-particle, spin-spin, and singlet diradical
 correlation functions which we now describe. In this section we shall
 be concerned with the  correlations of the electrons in 
 real-space. Thus in the following, indices $i$ and $j$ always refer to the
  (orthogonalised) $p_z$ \textit{atomic} orbitals on atoms $i$  and atom $j$
respectively. $\langle n_i^\sigma\rangle$ and $\langle S^z_i\rangle$  refer
to the average $\sigma$ occupancy and  $z$-component of the spin in these orbitals. 

\textit{Particle-particle}: 
\begin{equation}
C_\text{particle}(i, j) = 4 \left( \langle n_i^\alpha n_j^\beta \rangle -
  \langle n_i^\alpha \rangle \langle n_j^\beta\rangle \right)
\end{equation}
This measures the correlation between the  $\alpha$ population of 
orbital $i$ and $\beta$  population of orbital $j$. In
 a single-determinant wavefunction (such as the Kohn-Sham
 wavefunction) there are no  $\alpha\beta$ correlations and this quantity is
identically zero. 


\textit{Spin-spin}:
\begin{equation}
C_\text{spin}(i, j) = 4\left(\langle S^z_i S^z_j \rangle - \langle S^z_i \rangle \langle
  S^z_j\rangle\right) 
\end{equation}
This measures the correlation between the spin in
orbital $i$ and the spin in orbital $j$.  Note that
in wavefunctions that preserve the correct singlet-spin symmetry as used in this
work, $\langle S^z_i \rangle = \langle S^z_j \rangle = 0$. Because
there are $\alpha\alpha$ and $\beta\beta$ correlations from  the
Pauli principle even in  single determinant wavefunctions, this
quantity does not fully vanish in non-interacting systems. 


\textit{ Singlet diradical}:
\begin{align}
C_\text{diradical}(i, j) &= 2 \left( \langle d_i^\alpha d_j^\beta \rangle - \langle d_i^\alpha \rangle
  \langle d_j^\beta \rangle\right.  + \left.\langle d_i^\beta d_j^\alpha \rangle -
  \langle d_i^\beta \rangle 
   \langle d_j^\alpha \rangle \right)  \nonumber\\
d_i^\alpha& = n_i^\alpha (1-n_i^\beta) \label{eq:diradcorr}
\end{align}
The single occupancy operator $d_i^\alpha$ measures the probability that
an orbital $i$ is occupied with $\alpha$ spin without any simultaneous
$\beta$ occupancy. This and the joint
diradical probability density $\langle d_i^\alpha d_j^\beta \rangle$
were introduced by Dutoi \textit{et al.} \cite{Dutoi2004}.
 The function $C_\text{diradical}$ above
is obtained by removing  the  independent probabilities of single occupation (e.g. $\langle d_i^\alpha \rangle \langle
 d_j^\beta \rangle $) from the probability density of Dutoi \textit{et al.}, 
to give the correlation between single occupancies of orbitals $i$ and
orbital $j$ with opposite spin. Again,
because of Pauli type correlations in single determinant
wavefunctions, this quantity does not fully vanish in non-interacting systems.


To make  the meanings  of these correlation
functions explicit,  we can examine the following limiting
cases for 2-electron wavefunctions $\Psi$. (Here $\phi_1$ and $\phi_2$
are disjoint orthogonal atomic orbitals).
\begin{enumerate}
\item Singlet diradical $\Psi=\frac{1}{2} (\phi_1 \phi_2 + \phi_2 \phi_1)(\alpha \beta - \beta
\alpha)$. In this case, the above correlation functions assume their
extremum values of $1$ or $-1$. Thus  we would find $C_\text{particle}(1,
1)=C_\text{particle}(2, 2)=-1$, $C_\text{spin}(1,  1)=
C_\text{spin}(2,  2)=1$, $C_\text{diradical}(1,
1)=C_\text{diradical}(2, 2)=-1$, and $C_\text{particle}(1, 2)= 1$,
$C_\text{spin}(1,  2) = -1$, and $C_\text{diradical}(1, 2)=1$.
\item Closed-shell singlet  $\Psi=\frac{1}{2\sqrt{2}}
  (\phi_1 + \phi_2)(\phi_1 + \phi_2)(\alpha \beta - \beta \alpha)$. In
  this case all correlation functions are identically 0, reflecting
  the absence of $\alpha\beta$ correlation. 
\item Triplet $m_s=0$ diradical $\Psi=\frac{1}{2} (\phi_1 \phi_2 - \phi_2 \phi_1)(\alpha \beta + \beta
\alpha)$. Here all correlation function values are identical to those for the singlet
diradical wavefunction. (One observes that the $m_s=0$
triplet wavefunction differs from that of the singlet diradical only
in the phase relationship between orbital products such as
$\phi_1^\alpha \phi_2^\beta$ and $\phi_1^\beta \phi_2^\alpha$, thus
to distinguish the two one should examine averages such as
$\langle S^+_i S^-_j\rangle$. Note that $m_s=1$ triplet states
were used in this study).
\end{enumerate}

\subsection{Correlation functions}

\begin{figure*}

\centerline{\includegraphics*[width=5.3in]{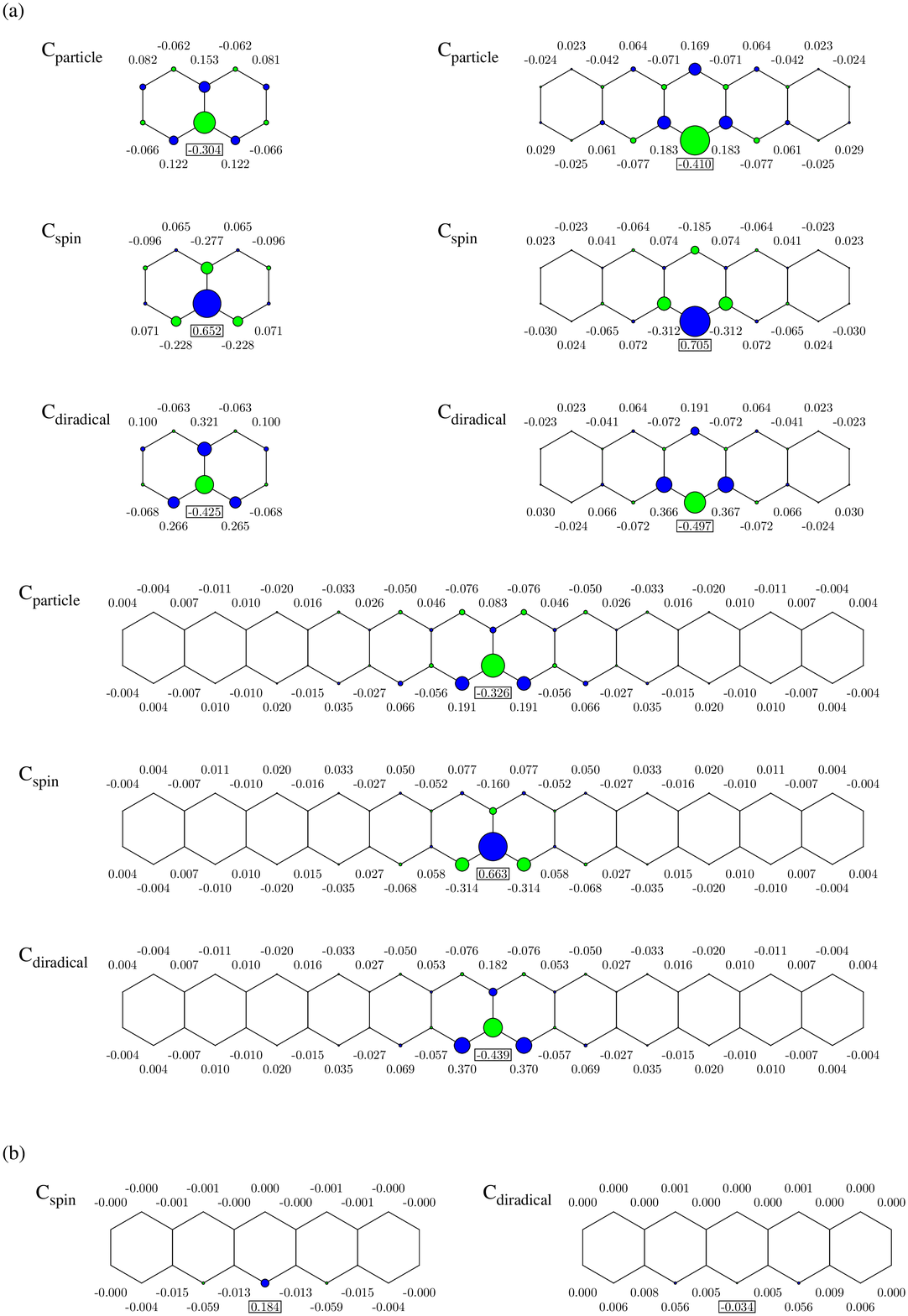}}

\caption 
{(a) Particle-particle,
spin-spin, and singlet diradical correlation functions evaluated for
napthalene, pentacene, and dodecacene in the STO-3G basis, fixing the reference point
at
the centre of the lower acene strand (indicated by the boxed value). The value of the correlation
function is indicated by the numbers; the size and colour of the
circles give the magnitude and sign of the correlation
function respectively. (b) Correlation functions in a non-interacting
model of pentacene.
}

\label{fig:allcorrs}
\end{figure*}

\begin{figure*}
\centerline{\includegraphics*[width=6.5in]{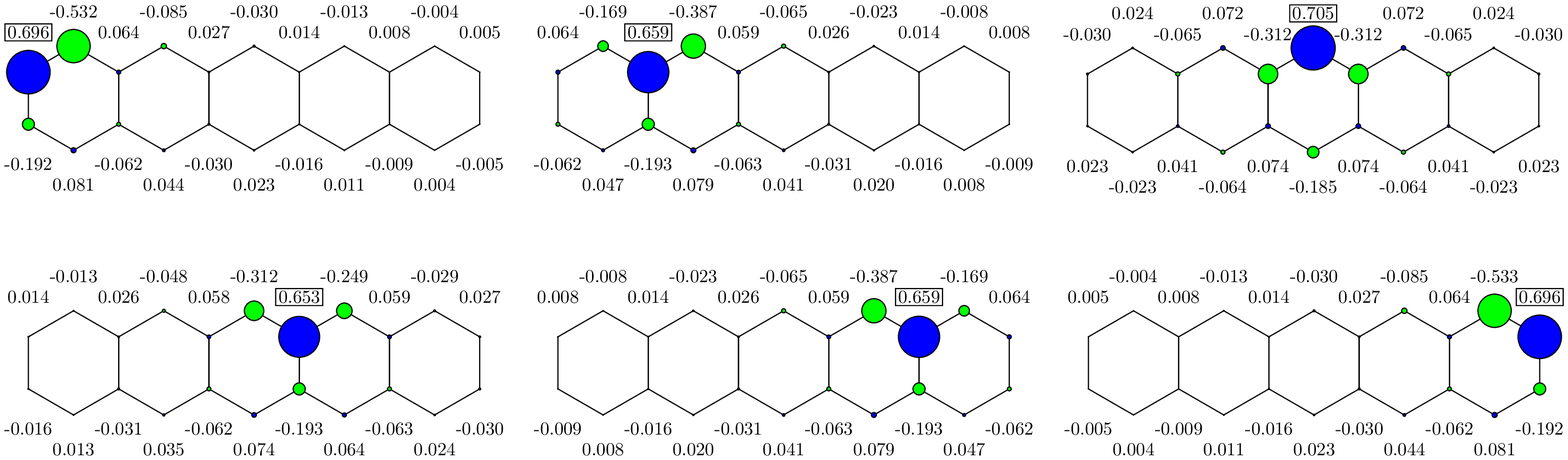}}

\caption 
{
Spin-spin correlation functions for pentacene as we move the
reference point of the correlation function (indicated by the boxed
value) around the ring.
}

\label{fig:pentacorrs}

\end{figure*}

Figs. \ref{fig:allcorrs} and \ref{fig:pentacorrs} show
plots of the different correlation functions evaluated for the singlet
ground states of napthalene,
pentacene and dodecacene. Since the correlation functions are functions of two positions, we
have plotted them as a function of the second position with the first
(reference) position fixed (indicated by the boxed value in the figures).

In Fig. \ref{fig:allcorrs} all plots have the reference  position
 fixed at the centre of the lower acene strand.  Examining
$C_\text{particle}$ we see that
it is large and negative  at the reference position. Thus given an $\alpha$
electron in this orbital, there is a significantly \textit{decreased} chance of
finding a simultaneous $\beta$ occupation of the orbital, or more simply,
double occupancy of the atomic orbital is disfavoured. Moving one atom
away,  $C_\text{particle}$ is large and positive reflecting an
\textit{increased} chance of finding the orbital to be occupied with
 opposite spin to that at the reference position. This antiferromagnetic
 correlation  continues further away  from the reference position in a 
pattern of positive and negative values of $C_\text{particle}$, though the
rapidly decreasing amplitudes indicate that the correlations are short-ranged. 

Examining the spin-spin $C_\text{spin}$ and singlet diradical
 $C_\text{diradical}$ correlation functions  yields a similarly consistent
picture. $C_\text{spin}$ is large and positive at the reference
position while $C_\text{diradical}$  is large and negative,  which
 both indicate that the orbital has a strong tendency towards 
single occupation.
The neighbouring atoms further show strong
single-occupancy, antiferromagnetic  correlation with large
negative (positive) values of $C_\text{spin}$ ($C_\text{diradical}$), and this correlation
 decreases rapidly further away from the reference position. 
While $C_\text{spin}$ and  $C_\text{diradical}$ do not identically
 vanish for a single determinant uncorrelated wavefunction, their
 corresponding plots  for pentacene in Fig. \ref{fig:allcorrs} show
 that aside from a small reduced propensity for double occupancy at the
 reference position which results from electron delocalisation, there
 are no significant antiferromagnetic correlations along the
 chains. (Recall that $C_\text{particle}$ is identically zero in the single
 determinantal wavefunction).

Comparing the  correlation functions of napthalene and
dodecacene shown in Fig. \ref{fig:allcorrs}, for which the reference atom is
in both cases at an ``inner'' position on the strand,  we see that
there is a (slight) increase in the
antiferromagnetic correlations  as the length of the
acene increases. In napthalene  the
correlation between the reference atom and the atom on the  neighbouring strand  is stronger
than the correlation to its neighbours on the same strand, a
situation which is reversed in the longer acenes. This is consistent
with the increasing difference between the ladder and rung C-C
bond-lengths, which leads to the view of the longer acenes as a pair
of coupled polyacetylene strands \cite{Houk2001, Bendikov2004}.


Fig. \ref{fig:pentacorrs} shows the spin-spin correlation plots where we move
the reference position around the ring. As expected the  
antiferromagnetic correlations persist as the reference position is
moved. Bond alternation is stronger near the edges of the pentacene
ring and this leads to  asymmetrical  correlations between
the reference position and its neighbours; stronger correlations are
observed across the shorter bonds.


\section{The nature of bonding in the acene polyradical state}

The correlation functions evaluated above  present
a dynamic picture of the electronic motion in the acenes. Tracking a
single electron as it makes its way around the ring, a second
electron is pulled along, antiferromagnetically coupled to the first
and distributed over the nearest neighbour atoms.
 
\begin{figure}
\centerline{\includegraphics*[width=3in]{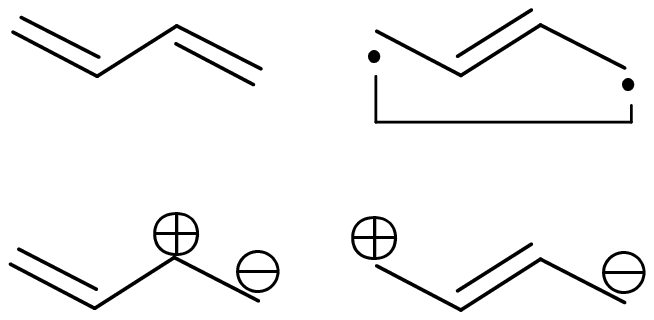}}
\caption{Covalent (top) and ionic (bottom) resonance structures for a conjugated
  system. Note that electron delocalisation requires a combination of both covalent
  and ionic resonance structures with roughly equal weights.}
\label{fig:resonance}

\end{figure}

Short-range antiferromagnetic correlations  naturally
bring to mind resonating valence bonds  \cite{Cooper2002,Goddard1978,Gerratt1997,Shaik2004,
GarciaBach1992, Gao2002}. Recall that we can expand any
wavefunction in terms of
resonance structures, which may be classified as covalent, singly ionic,
doubly ionic and so on (see Fig. \ref{fig:resonance}). (In this
language, the resonance structures are viewed only as a many-body basis for
expanding the wavefunction; the molecular geometry is
 fixed across the different structures). From our correlation
functions we see that the acene ground state is dominated by
\textit{covalent} resonance structures (no double occupancy of
the $p_z$ orbitals) with short-range spin-couplings (i.e. short-ranged
antiferromagnetic correlations). 

We should note that the use of the word resonance here  is different from the
colloquial usage where resonance structures are a simple metaphor
for delocalisation. In terms of the resonance structures as a physical
basis, electron delocalisation requires superposition of covalent \textit{and} ionic structures with roughly equal weights. 
In the limit where the wavefunction is comprised only of covalent structures, we
instead have an extreme \textit{localisation}: the electrons are
fixed and unpaired on each of the atomic sites with  only a spin  degree
of freedom, which fluctuates between different kinds of
spin-couplings. Wavefunctions which are predominantly 
covalent in nature   can  therefore be viewed as polyradicals as every
electron is unpaired in a localised, isoenergetic atomic orbital. 
The covalent nature of the  acene ground-state revealed by the
correlation functions  argues for this polyradical interpretation,
which is consistent with the picture given  earlier by the natural orbital occupations. 


Valence bond language in  conjugated $\pi$-systems has long been appreciated in the context of their
 low-lying states, which are conventionally
 classified as covalent or ionic depending on the main resonance
 structures.  Typically, covalent states appear more naturally described
 in the valence bond language. For example, in the
 polyenes, which may be thought of as making  the two legs of the
 acene ladder, the lowest   excitation  is a covalent $2A_g$ state which
 appears to have large \textit{double} excitation character from a molecular orbital
 viewpoint. However, this low-lying double excitation is easily
 understood in the  valence
bond language  as arising from the singlet-recoupling of two
singlet$\to$triplet excitations on adjacent double bonds
\cite{Schulten1972,Dunning1973,Ohmine1978,Hudson1982,Said1984,Wu2000}.
Valence bond descriptions and analyses have also been examined in the
context of radical electronic structure \cite{Salem1972, Wang2005}.
In traditional CAS calculations, to extract a
valence bond picture one usually re-parametrises the wavefunction
through a valence bond expansion (sometimes known as CASVB
 \cite{Hirao1996,Thorsteinsson1996,Thorsteinsson1997,Cooper1998,Thorsteinsson2000}). 
Such studies also find that the benzene ground-state should be
 viewed as a covalent state with antiferromagnetic spin-couplings, 
in accordance with what we have found for the acenes. The exponential size of the valence
bond basis limits the  CASVB analysis to small molecules, but as we have demonstrated, 
correlation functions provide an alternative mechanism to infer the
resonance nature of a state.

In a simple view of bonding, such as that afforded by the Hubbard \cite{Hubbard1963, Lieb1968} 
 or Pariser-Parr-Pople  models there are
two scales of energy, 
the resonance or hopping energy $t$ associated with delocalisation  and the  Coulomb repulsion energy $U$ associated with
double occupancy of an atomic orbital. When $U/t \gg 1$, we may be said to be in the
strongly interacting regime. Under such circumstances, the molecular
orbital picture begins to break down and instead the appropriate
qualitative wavefunction is the superposition of covalent resonance
structures as described above. A standard choice of parameters for conjugated
 polymers in  the Ohno parametrisation of the PPP Hamiltonian
 is $U=11.26$eV and $t=2.4$eV \cite{Ohno1964, Klopman1964, Raghu2002a}, placing
 systems such as the acenes
in the moderately strongly interacting spectrum of Hamiltonians and therein lies
  an understanding of the polyradical
character and covalent ground-state that we have observed.

\section{Conclusions}

In summary, motivated by predictions of unusual ground-states in the
longer acene molecules, we investigated 
acene electronic structure with high-level wavefunction theory.
Using  a new \textit{ab-initio} Density Matrix Renormalization Group
algorithm we could  carry out Complete Active Space
calculations  on the acenes from napthalene to dodecacene that correlated the full $\pi$-valence space. We
 find that the ground-state remains a singlet as the
chain-length increases, with a finite singlet-triplet gap in the
infinite chain limit. Detailed examination of the
wavefunctions, natural orbitals, and effective number of unpaired
electrons further reveals that the longer acenes exhibit singlet \textit{polyradical} 
character in their ground-state.
Through a series of correlation functions we observe that  electrons are
antiferromagnetically coupled in pairs on neighbouring atoms as they move around the acene chains.
These results are consistent with a view of the longer acenes as
moderately strongly interacting electronic systems, for which the
appropriate reference description is a polyradical wavefunction arising
from a resonance of predominantly  covalent valence bond structures. We
note that such a viewpoint is essential to understand the
excitations of these systems. Finally, our study illustrates that
 even  simple systems such as the acenes can provide unusual surprises
 in their electronic structure.



\begin{acknowledgments}
JH is funded by a Kekul\'{e} Fellowship of the Fond der Chemischen Industrie
(Fund of the German Chemical Industry). MA was supported by the Cornell Center
for Materials Research (CCMR) through their REU program. GKC
acknowledges support from
Cornell University, CCMR, the David
and
Lucile Packard Foundation in Science and Engineering, and the National Science
Foundation CAREER program CHE-0645380.
\end{acknowledgments}


\begin{thebibliography}{99}

\bibitem{Clar1964} E. Clar, \textit{Polycyclic Hydrocarbons} (Academic
Press, London, 1964).

\bibitem{Havey1997} R. G. Havey, \textit{Polycyclic Aromatic Hydrocarbons} (Wiley-VCH, New York, 1997).

\bibitem{Geerts1998} Y. Geerts, G. Kl\"{a}rner, and K. M\"{u}llen, in \textit{%
Electronic Materials: The Oligomer Approach}, edited by K. M\"{u}llen and G. Wagner (Wiley-VCH, Weinheim, 1998), p. 48.

\bibitem{Dimitrakopoulos2002} C. D. Dimitrakopoulos and P. R. L. Malenfant, Adv. Mater. \textbf{14}, 99 (2002).

\bibitem{Reese2004} C. Reese, M. Roberts, M. Ling, and Z. Bao, Mater. Today, 20 (2004).

\bibitem{BendikovRev} M. Bendikov, F. Wudl, and D. F. Perepichka, Chem. Rev. \textbf{104}, 4891 (2004).

\bibitem{Angliker1982} H. Angliker, E. Rommel, and J. Wirz, Chem.
Phys. Lett. \textbf{87}, 208 (1982).

\bibitem{Kertesz1983} M. Kertesz and R. Hoffmann, Solid State Comm. \textbf{47}, 97 
(1983).

\bibitem{Kivelson1983} S. Kivelson and O. L. Chapman, Phys. Rev. B \textbf{28}, 7236  
(1983).

\bibitem{Wiberg1997} K. B. Wiberg, J. Org. Chem. \textbf{62}, 5720 (1997).

\bibitem{Houk2001} K. N. Houk, P. S. Lee, and M. Nendel, J. Org. Chem. \textbf{66}, 5517 (2001).

\bibitem{Bendikov2004} M. Bendikov, H. M. Duong, K. Starkey, K. N. Houk, E. A. Carter, and F. Wudl, J. Am. Chem. Soc. \textbf{126}, 7416, 10493 (2004).

\bibitem{Mondal2006} R. Mondal, B. K. Shah, and D. C. Neckers, J. Am. Chem. Soc. \textbf{128}, 9612 (2006).

\bibitem{Bendikov2006} A. R. Reddy and M. Bendikov, Chem. Commun., 1179 (2006).

\bibitem{BallyBordenRev} T. Bally and W. T. Borden, Rev. Comp. Chem. \textbf{13}, 1 (1999).

\bibitem{StantonGaussRev} J. F. Stanton and J. Gauss, Adv. Chem. Phys. \textbf{125}, 101 (2003).

\bibitem{Slipchenko2002} L. V. Slipchenko and A. I. Krylov, J. Chem. Phys. \textbf{117}, 4694 (2002).

\bibitem{Salem1972} L. Salem, C. Rowland, Angew. Chem. Int. Ed. \textbf{11}, 92 (1972).

\bibitem{Borden1977} W. T. Borden, E. R. Davidson, J. Am. Chem. Soc. \textbf{99}, 4587 (1977).

\bibitem{BordenBook} W. T. Borden, \textit{Diradicals} (Wiley, New York, 1982).

\bibitem{Rajca1994} A. Rajca, Chem. Rev. \textbf{94}, 871 (1994).

\bibitem{Jung2003} Y. Jung and M. Head-Gordon, ChemPhysChem \textbf{4}, 522 (2003).

\bibitem{Krylov2005} A. I. Krylov, J. Phys. Chem. A \textbf{109}, 10638 (2005).

\bibitem{Roos1987} B. O. Roos, Adv. Chem. Phys. \textbf{69}, 399 (1987).

\bibitem{Kawashima1999} Y. Kawashima, T. Hashimoto, H. Nakano, and K. Hirao, Theor. Chem. Acc. \textbf{102}, 49 (1999).

\bibitem{White1992} S. R. White, Phys. Rev. Lett. \textbf{69}, 2863 (1992).

\bibitem{White1993} S. R. White, Phys. Rev. B \textbf{48}, 10345 (1993).

\bibitem{WhiteMartin1999} S. R. White and R. L. Martin, J. Chem. Phys. \textbf{110}, 4127 (1999).

\bibitem{Raghu2002a} C. Raghu, Y. Anusooya Pati, and S. Ramasesha, Phys. Rev. B \textbf{65}, 155204 (2002).

\bibitem{Raghu2002b} C. Raghu, Y. Anusooya Pati, and S. Ramasesha, Phys. Rev. B \textbf{66}, 035116 (2002).

\bibitem{Hachmann2006} J. Hachmann, W. Cardoen, and G. K.-L. Chan, J. Chem. Phys. \textbf{125}, 144101 (2006).

\bibitem{Chan2002} G. K.-L. Chan and M. Head-Gordon, J. Chem. Phys. \textbf{116}, 4462 (2002).

\bibitem{Chan2004} G. K.-L. Chan, J. Chem. Phys. \textbf{120}, 3172 (2004).

\bibitem{LYP1988} C. Lee, W. Yang, and R. G. Parr, Phys. Rev. B \textbf{37}, 785 (1988). 

\bibitem{Becke1993} A. D. Becke, J. Chem. Phys. \textbf{98}, 5648 (1993).

\bibitem{sto3g} W. J. Hehre, R. F. Stewart, and J. A. Pople, J. Chem. Phys. \textbf{51}, 2657 (1969).

\bibitem{dz1} T. H. Dunning Jr., J. Chem. Phys. \textbf{53}, 2823 (1970).

\bibitem{dz2} T. H. Dunning Jr. and P. J. Hay, in \textit{Methods of Electronic
Structure Theory}, Vol. 2, edited by H. F. Schaefer III (Plenum Press, New York, 1977). 

\bibitem{g03} Gaussian 03, Revision C.02, M. J. Frisch \textit{et al.}, Gaussian, Inc., Wallingford CT, 2004.

\bibitem{BirksBook} J. B. Birks, \textit{Photophysics of aromatic molecules} (Wiley, London, 1970).

\bibitem{Schiedt1997} J. Schiedt and R. Weinkauf, Chem. Phys. Lett. \textbf{266}, 201 (1997).

\bibitem{Sabbatini1982} N. Sabbatini, M. T. Indelli, M. T. Gandolfi, and V. Balzani, J. Phys. Chem. \textbf{86}, 3585 (1982).

\bibitem{Burgos1977} J. Burgos, M. Pope, Ch. E. Swenberg, and R. R. Alfano, Phys. Status Solid. B \textbf{83}, 249 (1977).

\bibitem{Andersson1992} K. Andersson and P.-{\AA}. Malmqvist and
  B. O. Roos,
J. Chem. Phys. \textbf{96}, 1218 (1992).

\bibitem{Hirao1992} K. Hirao, Chem. Phys. Lett. \textbf{190}, 374 (1992).

\bibitem{PPP1} R. Pariser and R. G. Parr, J. Chem. Phys. \textbf{21}, 466, 767 (1953). 

\bibitem{PPP2} J. A. Pople, Trans. Faraday Soc. \textbf{49}, 1375 (1953).

\bibitem{Doehnert1980} D. D\"{o}hnert and J. Kouteck\'{y}, J. Am. Chem. Soc. \textbf{102}, 1789 (1980).

\bibitem{Takatsuka1978a} K. Takatsuka, T. Fueno, and K. Yamaguchi, Theor. Chim. Acta \textbf{48}, 175 (1978). 

\bibitem{Takatsuka1978b} K. Takatsuka and T. Fueno, J. Chem. Phys. \textbf{69}, 661 (1978).

\bibitem{Bochicchio1998} R. C. Bochicchio, J. Mol. Struct. THEOCHEM \textbf{429}, 229 (1998).

\bibitem{Staroverov2000a} V. N. Staroverov and E. R. Davidson, J. Am. Chem. Soc. \textbf{122}, 186 (2000).

\bibitem{Staroverov2000b} V. N. Staroverov and E. R. Davidson, Chem. Phys. Lett. \textbf{330}, 161 (2000).

\bibitem{HeadGordon2003a} M. Head-Gordon, Chem. Phys. Lett. \textbf{372}, 508 (2003). 

\bibitem{Bochicchio2003} R. C. Bochicchio, A. Torre, and L. Lain, Chem. Phys. Lett. \textbf{380}, 486 (2003).

\bibitem{HeadGordon2003b} M. Head-Gordon, Chem. Phys. Lett. \textbf{380}, 488 (2003).

\bibitem{Fano1998} G. Fano, F. Ortolani, and L. Ziosi, J. Chem. Phys. \textbf{108},
9246 (1998).

\bibitem{Dutoi2004} A. D. Dutoi, Y. Jung, and M. Head-Gordon, J. Phys. Chem. A \textbf{108}, 10270 (2004).

\bibitem{Cooper2002} D. Cooper, \textit{Valence Bond Theory} (Elsevier, Amsterdam, 2002).

\bibitem{Goddard1978} W. A. Goddard III and L. B. Harding, Ann. Rev. Phys. Chem. \textbf{29}, 363 (1978).

\bibitem{Gerratt1997} J. Gerratt, D. L. Cooper, P. B. Karadakov, and M. Raimondi, Chem. Soc. Rev. \textbf{26},
87 (1997).

\bibitem{Shaik2004} S. Shaik and P. C. Hiberty, Rev. Comp. Chem. \textbf{20}, 1 (2004).

\bibitem{GarciaBach1992} M. A. Garcia-Bach, A. Pe\~{n}aranda, and D. J. Klein, Phys. Rev. B \textbf{45}, 10891 (1992).

\bibitem{Gao2002} Y. Gao, C.-G. Liu, and Y.-S. Jiang, J. Phys. Chem. A \textbf{106}, 2592 (2002).

\bibitem{Schulten1972} K. Schulten and M. Karplus, Chem. Phys. Lett. \textbf{14}, 305 (1972).

\bibitem{Dunning1973} T. H. Dunning, R. P. Hosteny, and I. Shavitt, J. Am. Chem. Soc. \textbf{95}, 5067 (1973).

\bibitem{Ohmine1978} I. Ohmine, M. Karplus, and K. Schulten, J. Chem. Phys. \textbf{68}, 2298 (1978).

\bibitem{Hudson1982} B. S. Hudson, B. E. Kohler, K. Schulten, in \textit{Excited
States}, Vol. 6, edited by E. C. Lim (Academic Press, New York, 1982), p. 1.

\bibitem{Said1984} M. Said, D. Maynau, and J. P. Malrieu, J. Am. Chem. Soc. \textbf{106}, 580 (1984).

\bibitem{Wu2000} W. Wu, D. Danovich, A. Shurki, and S. Shaik, J. Phys. Chem. A \textbf{104}, 874 (2000).

\bibitem{Wang2005} T. Wang and A. I. Krylov,
  J. Chem. Phys. \textbf{123}, 104304 (2005).

\bibitem{Hirao1996} K. Hirao, H. Nakano, K. Nakayama, and M. Dupuis, J. Chem. Phys. \textbf{105}, 9227 (1996).

\bibitem{Thorsteinsson1996} T. Thorsteinsson, D. L. Cooper, J. Gerratt, P. B. Karadakov, and M. Raimondi, Theor. Chim. Acta \textbf{93}, 343 (1996).

\bibitem{Thorsteinsson1997} T. Thorsteinsson, D. L. Cooper, J. Gerratt, and M. Raimondi, in \textit{Quantum Systems in Chemistry and Physics: Trends in Methods and
Applications: A new approach to valence bond calculations: CASVB}, edited by R. McWeeny, J. Maruani, Y. G. Smeyers, and S. Wilson (Kluwer, Dordrecht, 1997). 

\bibitem{Cooper1998} D. L. Cooper, T. Thorsteinsson, and J. Gerratt, Adv. Quant. Chem. \textbf{32}, 51 (1998). 

\bibitem{Thorsteinsson2000} T. Thorsteinsson and D. L. Cooper, in \textit{Quantum
Systems in Chemistry and Physics}, Vol. 1: \textit{Basic problems and models systems: An
overview of the CASVB approach to modern valence bond calculations}, edited by
A. Hern\'{a}ndez-Laguna, J. Maruani, R. McWeeny, and S. Wilson (Kluwer,
Dordrecht, 2000), p. 303. 

\bibitem{Hubbard1963} J. Hubbard, Proc. Roy. Soc. Lond. A \textbf{276}, 238 (1963).

\bibitem{Lieb1968} E. H. Lieb and F. Y. Wu, Phys. Rev. Lett. \textbf{20}, 1445 (1968).

\bibitem{Ohno1964} K. Ohno, Theor. Chim. Acta \textbf{2}, 219 (1964). 

\bibitem{Klopman1964} G. Klopman, J. Am. Chem. Soc. \textbf{86}, 4550 (1964).


\bibitem{supportingmaterial} See EPAPS Document No. for further details of the calculations, energies, optimized geometries for all calculated oligoacenes at UB3LYP/6-31G(d) level, the data for Fig. \ref{fig:measures}, and complete Ref. \cite{g03}. This document can be reached through a direct link in the online article's HTML reference section or via the EPAPS homepage (http://www.aip.org/pubservs/epaps.html).



\end{thebibliography}
\end{document}